\documentclass[a4paper]{article}

\usepackage{amsfonts, amsmath, amssymb, amsbsy, graphicx}
\usepackage[utf8]{inputenc}
\usepackage[english, ngerman]{babel}
\usepackage[T1]{fontenc}
\usepackage{wrapfig}

\begin{document}
\selectlanguage{english}

\begin{center}
\textbf{\LARGE Prospect of Plate Archive Photometric Calibration by GAIA SED Fluxes}
\end{center}

\begin{center}
\textbf{Maryam Raouph$^1$, Andreas Schrimpf$^1$, Peter Kroll$^2$}
\end{center}

\begin{center}
{\it
\noindent $^1$History of Astronomy and Observational Astronomy, Physics Department, Philipps University Marburg \\
$^2$Sonneberg Observatory, Sonneberg, Germany }
\end{center}

\begin{abstract}
This study aims to improve the photometric calibration of astronomical photo plates. The Sonneberg Observatory's sky patrol was selected, comprising about 300,000 plates, and the digitization workflow is implemented using PyPlate. The challenge is to remove zero point offsets resulting from differences in color sensitivity in the photo plates' emulsion response. By utilizing the Gaia DR3 dataset and the GaiaXPy tool, we are able to obtain a consistent astrometric and photometric calibration of the Sonneberg plates and those of other archives such as APPLAUSE.
\\
\\
\noindent Poster presented at the Annual meeting of the German Astronomical Society, Berlin (2023) 
\end{abstract}

\section{Introduction}

Photographic plates were widely used in professional astronomy in the 20$^\text{th}$ century to record information on celestial objects and have since been digitized to a significant extent. They work by diffusing photosensitive chemicals into a gel to create an emulsion on a flat glass surface that reacts to light exposure. Sonneberg photo plate archive is located at Thuringia, Germany and its plates taken between 1923 and 2010 provide valuable information for studying long-term changes in variable´objects as a fast-growing field, called time-domain.

\begin{figure*}[htp!]
	\centering

	\includegraphics[width=1.0\textwidth]{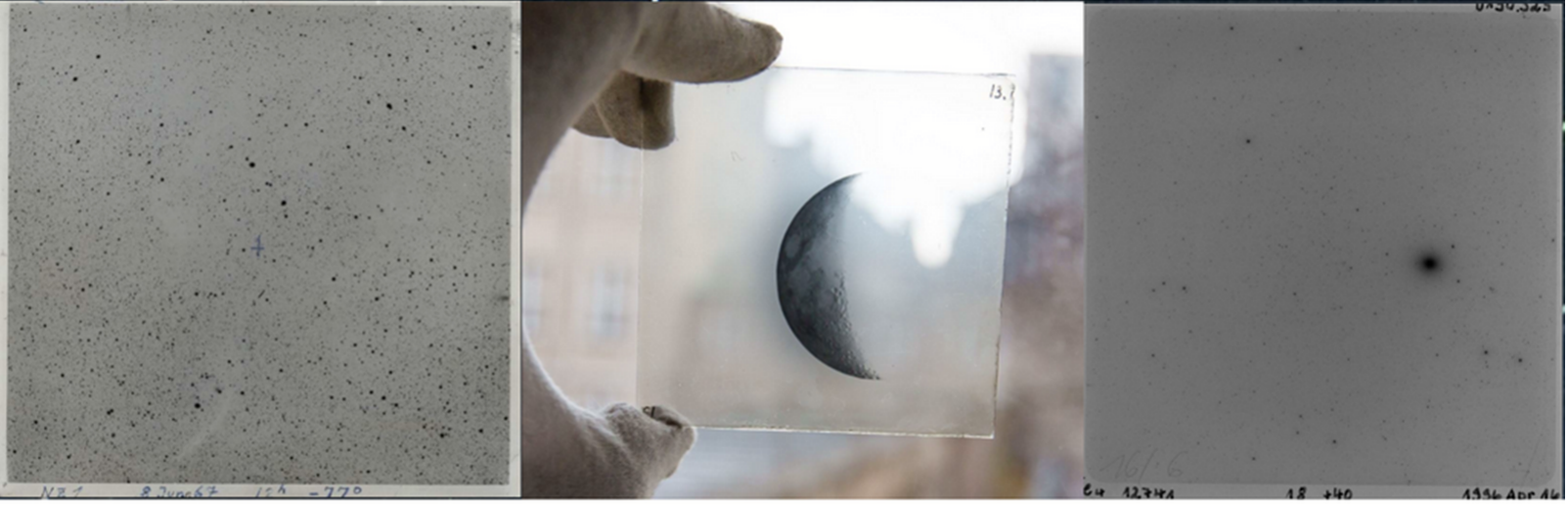}

	\caption{Left: The Chamaeleon constellation in the southern sky (Ref: Dr. Karl-Remeis-Observatory, Bamberg).  Middle:  Glass plates from  1909-1922 show the Moon in different phases (Ref: Østervold Observatory, Niels Bohr Institute). Right: Vega, the brightest star of the northern constellation of Lyra, 1996 (Ref: Sonneberg Observatory, Germany)}
	\label{fig1_raouph_photoplates}
\end{figure*}

Long-term photometry studies almost always lack homogeneity and comparability of the data because of
the reason of using different emulsions in series of observations. The difference in the color response of the emulsions makes a decisive change in the processing method of the plate in terms of color sensitivity \cite{froehlich2002}. For instance, such changes might make small offsets in zero points, on the level of $< 0.1$ mag, and are very hard to determine and remove from the time series \cite{hippke2017}.

\newpage
\section{Strategies and Methods}

\begin{figure}[htp!] % {r}{0.6\textwidth}
	\centering
	\includegraphics[width=0.60\textwidth]{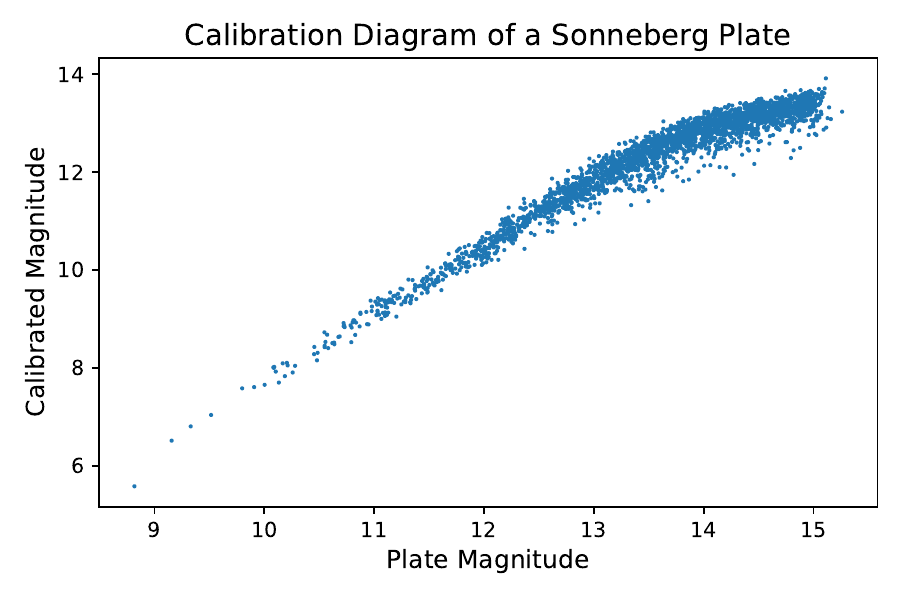}
	\caption{The calibrated photometric magnitudes versus the instrumental magnitudes in annular bin 1. Plate no. 9606, Sonneberg Sky Patrol, blue sensitive, 16.9.1982, centered at RA 1 hour, DEC 20 degree.  }
	\label{fig2_raouph_calib}
\end{figure}

Carrying out tasks in the digitization process of Sonneberg photo plates contains the extraction of
sources on direct images, astrometric and photometric calibration is implemented by an open-source
python package called PyPlate \cite{pyplate}. In figure \ref{fig2_raouph_calib} the calibration of a blue sensitive plate from the Sonneberg Sky Patrol campaign using PyPlate 4 is shown. The nonlinearity of the photometric sensitivity of the plates clearly can be identified.

\subsection{Color Term Correction: A Simple Solution}

The color term method is a widely used approach in photometry for avoiding plate offset by calibrating instrumental magnitudes to a standard system. The color term of the plate, defining the color response of the emulsion and filter as proposed by DASCH (Digital Access to a Sky Century at Harvard) \cite{dasch}  is defined as:

\begin{equation}
m_{\mathrm{cat}} = V_{\mathrm{cat}} + c(B_{\mathrm{cat}} - V_{\mathrm{cat}})
\label{dscheqn}
\end{equation}

\noindent where $m_{\mathrm{cat}} $ is the reference catalog magnitude transformed in the plate system, $ B_{\mathrm{cat}} $ and $ V_{\mathrm{cat}} $ the reference Johnson $B$ and $V $ magnitude, respectively, and $ c $ is the color term. PyPlate 3 uses UCAC4 and APASS as reference catalogs, and PyPlate 4 Gaia EDR3.

The tight correlation shown in fig. \ref{fig2_raouph_calib} indicates installing and using Pyplate was successful, and it found the best color term by minimization of the scatter for the entire plate.

\begin{figure*}[htp!]
	\centering

	\includegraphics[width=0.60\textwidth]{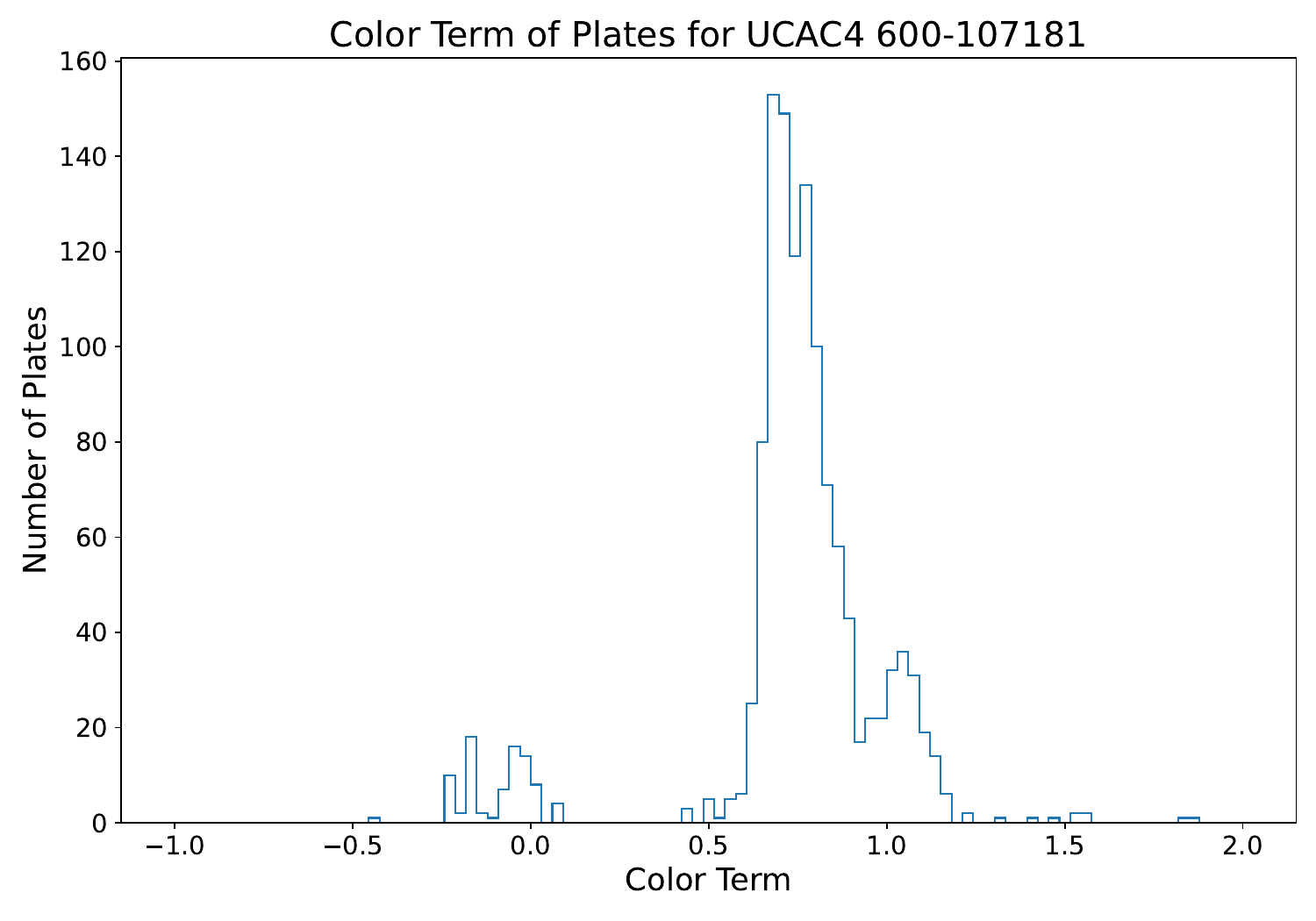}

	\caption{Color terms extracted from a series of plates from the APPLAUSE DR3 \cite{applause} containing the constant star UCAC4 600-107181. The coordinates is RA 300.2134071 and Dec +29.8457770 (ICRS).}
	\label{fig3_raouph_color_terms}
\end{figure*}

However, the color term has limitations and potential sources of error that should be carefully considered. In order to evaluate the color term approach we picked a constant star in a series of plates from the APPLAUSE DR3. This series contains blue and yellow sensitive plates and fig. \ref{fig3_raouph_color_terms} reveals three distributions: a color term of about 0 results from yellow plates (almost matching Johnson V), a color term of 1 comes from blue plates, matching Johnson B. Most of the plates in this series belong to a distribution off $c$ centered at 0.7, indicating an intermediate spectral sensitivity of the plates.

\begin{figure*}[htp!]
	\centering

	\includegraphics[width=0.64\textwidth]{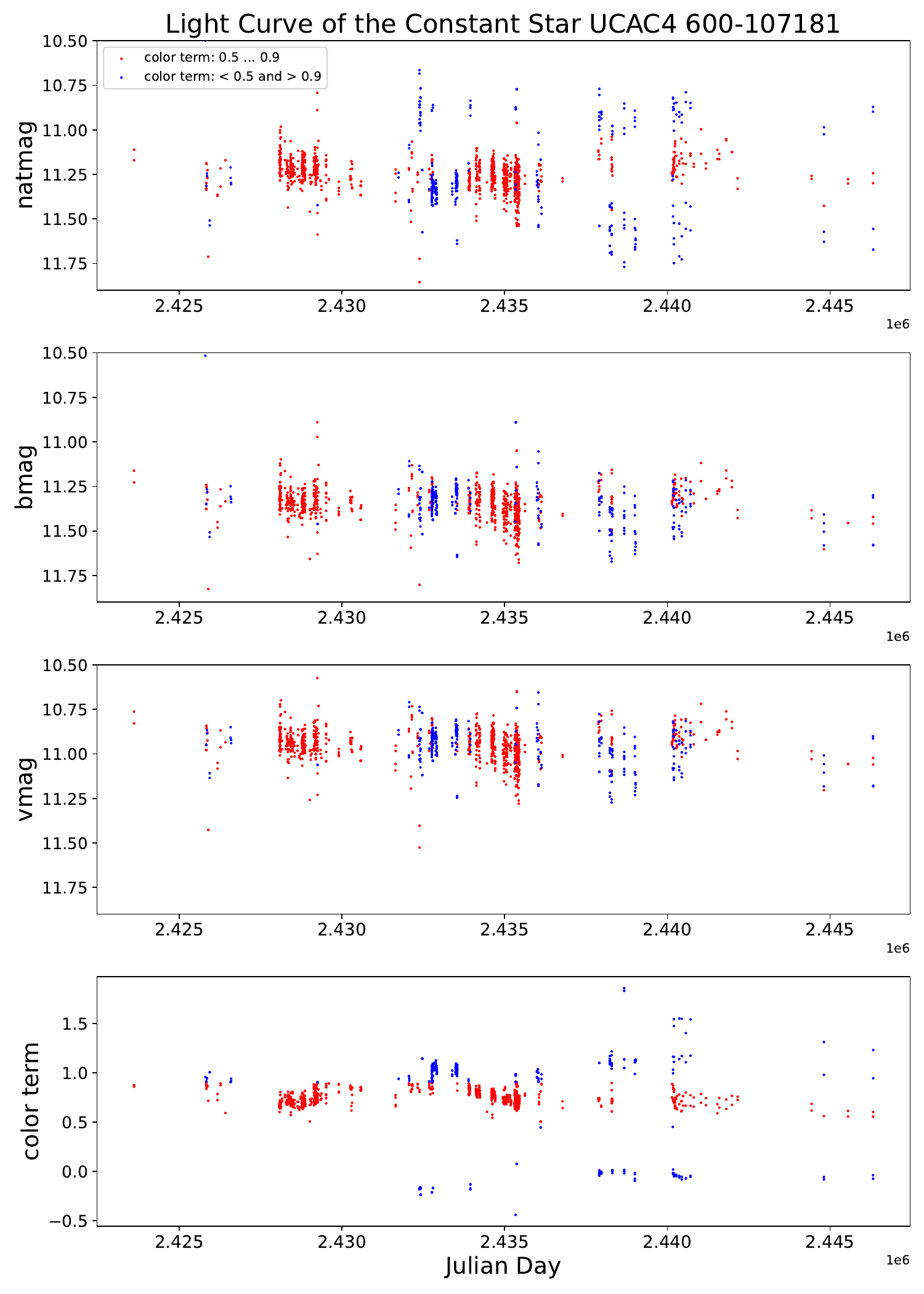}

	\caption{Light curve of the constant star UCAC4 600-107181 from APPLAUSE DR3. Upper panel: calibrated magnitude of the star in the plate's natural photometric systems, varying with the color response of the plates. Second and third panels: Calibrated magnitude of the star in Johnson B and V passbands. Lower panel: Color terms of the plates.}
	\label{fig4_raouph_lightcurve}
\end{figure*}

In fig. \ref{fig4_raouph_lightcurve} the  the light curve of the constant star 600-107181 obtained over 62 years of photometry using 616 photo plates from Applause DR3 is plotted.

The large variations of the light curve are correlated with color terms outside of the
main peak in fig \ref{fig3_raouph_color_terms}. And, the trends of the B and V light curves follow the varying color term even within the peak around 0.7. 

\newpage
Besides, the transition from APPLAUSE Database Release 3 (DR3) to the new DR4 release introduces a notable shift in the color term parameter range, a change closely linked to the use of the Gaia EDR3 catalog as a reference. This shift reflects the challenge of adapting to the broader spectral
information provided by GAIA, as opposed to the other reference catalogs.

\subsection{Synthetic Photometry: The Advanced solution}

Using a linear correction of a color index to find the magnitudes of the plates’ internal photometric systems is not appropriate!.

\emph{
The proposed solution to more advanced photometric calibration in this research is using the SEDs of
the full Gaia data release 3 (DR3) that has been released on 13 June 2022.
}

\begin{figure*}[htp!]
	\centering
	\includegraphics[width=0.45\textwidth]{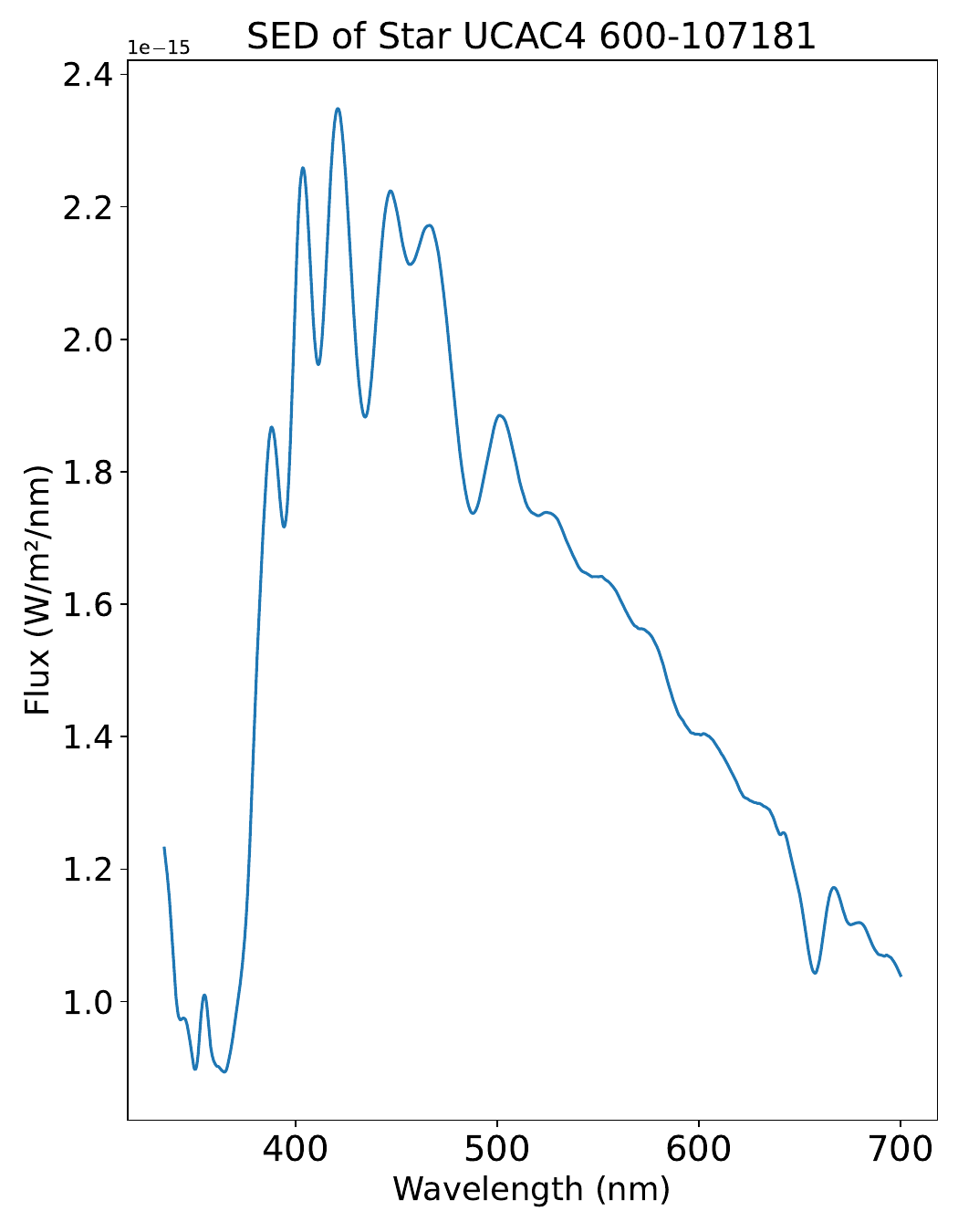}
	\includegraphics[width=0.45\textwidth]{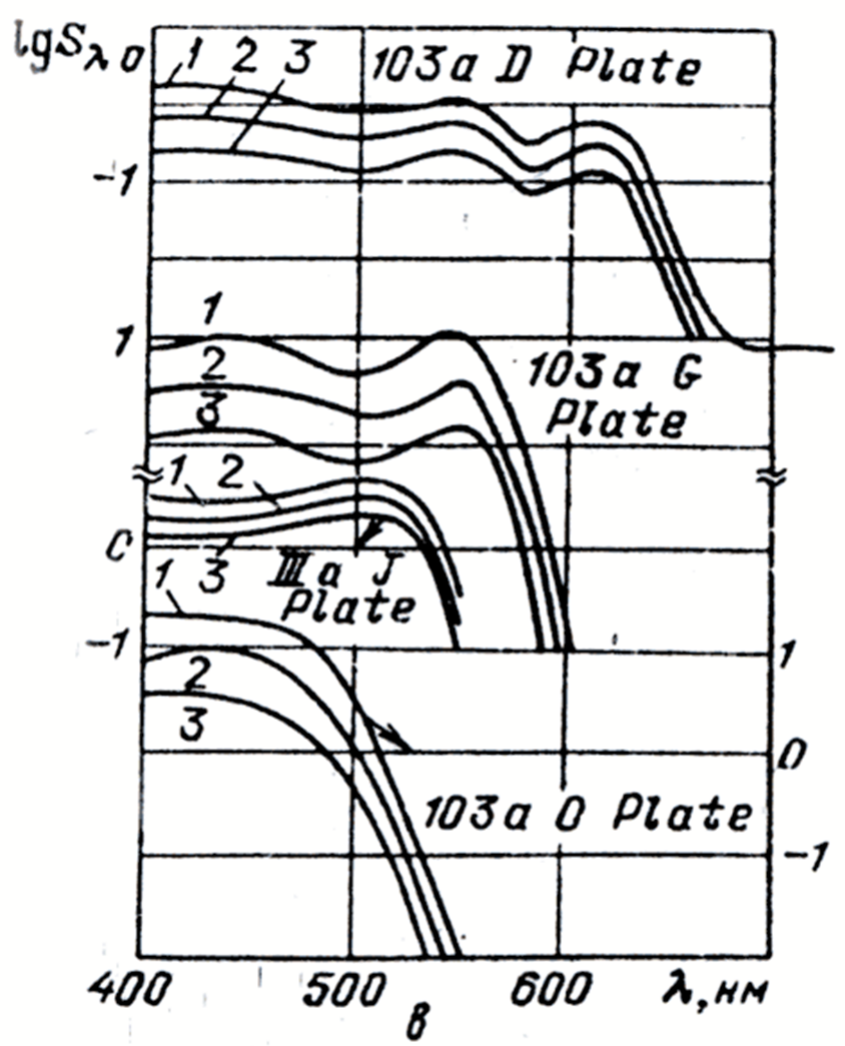}
	\caption{Left: SED of star 600-107181 retrieved from Gaia DR3 and processed with GaiaXPy python library \cite{deangeli2023}. Right: Spectral sensitivity of some Kodak emulsions used for the photo plates of the applause archive.}
	\label{fig5_raouph_sed}
\end{figure*}
The plates used in this analysis are Kodak 103a-D and Kodak 103a-O. The ''O'' plates have low sensitivity in the V band. 

The Gaia Data Release 3 includes low-resolution spectrophotometry for over 220 million sources
covering a wavelength range of 330 nm to 1050 nm (XP spectra). Using these spectral information
now for the first time it is possible to calculate the mean flux in any photometric system within the given spectral range by :

\begin{equation}
\displaystyle  	\left\langle f_\lambda \right\rangle = \frac{\int f_\lambda (\lambda) S(\lambda) \lambda\, d\lambda}{\int  S(\lambda) \lambda\, d\lambda}
\end{equation}
\noindent
where $ S(\lambda) $  is the transmission curve of a plate and $f_\lambda (\lambda)$ the spectral energy distribution of a star. This synthetic flux has to be converted into a magnitude, which then is used for calibration of the instrumental magnitudes of stars of the plate \cite{montegriffo2023}. To extract a light curve of a single star from many plates with differing
emulsions, the magnitudes of the star in each of the plates have to be back-transformed to a common
photometric system using the above formula backwards.

\section{Discussion}

The latest APPLAUSE data release 4 (DR4) benefits from the astrometric and photometric calibration
using Gaia early data release 3 (EDR3), resulting in better sky coverage and accuracy. However, due to the mismatch of filters used with Gaia, the photometric calibration using this catalog is not improved. The high positional resolution of Gaia data also reveals an issue with single sources being resolved as multiple stars, making photometric calibration impossible. By applying the proposed solution, we dispose of these drawbacks and resolve the problems in combining photometric results of photo plates with differing emulsions. That will obtain consistent astrometric and, for the first time, photometric results from different archives and ultimately allow the combining of data from every existing astronomical photo plate archive.


\begin{thebibliography}{20}
\bibitem{froehlich2002} Fröhlich, H.-E., Tschäpe, R., Rüdiger, G., and Strassmeier, K. G., EK Draconis: Long-term photometry on Sonneberg Sky-Patrol plates, A\&A 391, 659–663 (2002).
\bibitem{hippke2017} Hippke, M., et al., Sonneberg Plate Photometry for Boyajian's Star in Two Passbands, ApJ, 837, 85 (2017).
\bibitem{pyplate} Tuvikene, T., the APPLAUSE Collaboration, PyPlate: a software package for processing digitized astronomical photographic plates, Large Surveys with Small Telescopes, Bamberg 
(2019).
\bibitem{dasch} Laycock, S., Tang, S., Grindlay, J., Los, E., Simcoe, R. and Mink, D., Digital Access to a Sky  Century at Harvard: Initial Photometry and Astrometry, AJ 140, 1062 (2010).
\bibitem{applause} APPLAUSE DR3: https://www.plate-archive.org/cms/documentation/dr3/.
\bibitem{deangeli2023} De Angeli F, et al., Gaia Data Release 3: Processing and validation of BP/RP
low-resolution spectral data, A\&A 674, A2 (2023).
\bibitem{montegriffo2023} Montegriffo P, Bellazini, M., and the Gaia Collaboration, Gaia Data Release 3: The Galaxy in your preferred colours. Synthetic photometry from Gaia low-resolution spectra, A\&A 674, A33 (2023).
\end{thebibliography}
\end{document}